\documentstyle[12 pt, epsfig]{article}

\def\slash#1{\,/\kern-7pt#1}
\def\rd{\partial}

\def\darr#1{\raise1.5ex\hbox{$\leftrightarrow$}\mkern-16.5mu #1}

\def\rds{/\kern-6pt\rd}

\newcommand{\be}{\begin{equation}} \newcommand{\bea}{\begin{eqnarray}}
\newcommand{\ee}{\end{equation}} \newcommand{\eea}{\end{eqnarray}}
\newcommand{\ba}[1]{\left(\begin{array}{#1}}
\newcommand{\ea}{\end{array}\right)}

\newcommand{\nn}{\nonumber}

\newcommand{\id}{\kern0.2em\rule{0.1mm}{0.71em}
\kern0.12em\rule{0.1mm}{0.71em} \kern-
0.27em\rule[0.68em]{0.27em}{0.1mm} \kern-
0.30em\rule{0.44em}{0.1mm}\rule{0.1em}{-1mm}}

\begin{document}
\begin{titlepage}
\begin{center}
\hfill ITP/NSF-98-064, UCB/PTH-98/27, LBNL-41850

\hfill	hep-th/9805129

\vskip 1.5 cm

{\large \bf Aspects of Large N Gauge Theory Dynamics \\
as Seen by String Theory  }

\vskip 1.5 cm

{\large David J. Gross$^1$ and Hirosi Ooguri$^{1,2,3}$}\\

\vskip 1cm

{$^1$ Institute for Theoretical Physics, University of California, \\
		Santa Barbara, CA 93106 \\
$^2$ Department of Physics, University of California, \\
	Berkeley, CA 94720\\
$^3$ Theory Group, Lawrence Berkeley National Laboratory, \\
			Berkeley, CA 94720}
%\date{May, 1998}
\end{center}

\vskip	.5cm

\begin{abstract}
In this paper we explore some of the features of large $N$ supersymmetric and
nonsupersymmetric gauge
theories using Maldacena's duality conjectures. We shall show that the
resulting strong
coupling behavior of the gauge theories
is consistent with our qualitative expectations of
these theories.  Some of these consistency checks are highly nontrivial and
give additional evidence for the validity of the proposed dualities.

\end{abstract}

\end{titlepage}

\newpage

\section{Introduction}

The newest, and perhaps  most interesting, of the dualities of string
theory is that conjectured
by Maldacena, which relates the large $N$ expansion of conformal field
theory in $d$
dimensions to string theory in a $AdS_{d+1} \times M$
spacetime background (where $AdS_{d+1}$ is  $(d+1)$-dimensional
Anti de-Sitter space and $M$ is a compact space) \cite{malda1}.
The dictionary that relates these dual descriptions
identifies the $1/N$ expansion of  the field theory to the perturbative
expansion of the
string theory, and the  strong coupling expansion of the field theory to
the $\alpha'$
expansion of the string theory.
This conjecture offers the exciting possibility of using perturbative
string theory to
explore the large $N$ limit of field theory.

The simplest case of Maldacena's
conjecture is the duality between large $N$
supersymmetric, conformally invariant, $SU(N)$
gauge theory in four dimensions
(with coupling $g^2_{YM}$)  and type IIB string theory  expanded about an
$AdS_5\times {\bf S}^5$ background.
Here the string coupling, $g_{st}, $ is proportional to
 $g^2_{YM}$;  $N$ equals, in  the string theory,  the magnitude
of the five-form flux on the five-sphere; and  $(g^2_{{YM}}N)^{1/4}$
is proportional to the  radius of curvature of the background
$AdS_{5}$ space. One can therefore hope
to calculate gauge theory correlation functions, for large $N$ and large
$(\lambda = g^2_{YM}N)$,  in terms of 
weak coupling string theory in the semiclassical
approximation --- {\it{i.e.}}  supergravity.

The precise relation between the gauge theory correlation functions and the
supergravity
effective  action has been given by \cite{gkp,w1}, following the earlier
works \cite{earlier}.
In particular  this prescription determines the dimensions of operators in
the conformal field
theory
in terms of the masses of particle in the string theory. This
correspondence has been
checked for the duality between $SU(N)$ gauge theory in four dimensions
 and type IIB string theory  expanded about an
$AdS_{5}\times {\bf S}^5$ background; where it was shown that there is a
precise
correspondence between the chiral fields of the conformal gauge theory and
the finite mass
string states in the above limit, including the complete infinite tower
of massive Kaluza-Klein states of ten-dimensional supergravity on the
5-sphere \cite{ho,f}
%. This
%correspondence allows one to determine the anomalous dimension of the 
%chiraloperators
%in the strong coupling ($\lambda \to \infty $) limit of gauge theory.

In \cite{malda2,ry},
it was shown that the strong coupling limit of the large
Wilson loop for large $N$  can be  evaluated   using semiclassical string
theory, thereby obtaining the interaction energy between infinitely massive
quarks and anti-
quarks (external sources in the fundamental representation of
$SU(N)$),  separated by distance $R$
as
\be
E_{q\bar q}=-{4 \pi^2 \over {\Gamma (1/4)}^4}{\sqrt{2\lambda} \over R},
\ee
a result that is completely consistent with our limited understanding of
the gauge theory,
wherein the $1/R$ behavior is dictated by conformal invariance. The
proportionality to $\sqrt{\lambda}$ suggests that the
Coulomb forces is somewhat reduced from the weakly coupled
value of $\lambda$.
Similar calculations have been performed for the monopole-monopole and
monopole-quark potential, yielding,  as expected, $S$-dual  expressions
\cite{mina}.

One can regard   Maldacena's duality as realizing the long sought goal of
finding the master
field representation of large $N$ gauge theory correlation functions.
What is most surprising from this point of view is that the master field
lives in a compactified
ten-dimensional space-time, and corresponds to supersymmetric
type IIB string theory. That there should exist a string representation of
the ${\cal N}=4$
conformally invariant
large $N$ gauge  theory is somewhat surprising, since the traditional
arguments  for such a
representation have been for confining theories, whereas here we have a
string theory
for the    Coulomb phase of the gauge theory. Thus, even though  the Wilson
loop is given by
the minimal area classical string configuration spanning the loop, the fact
that the loop can
meander into the extra dimensions and the nature of the
geometry
of $AdS$ space lead to a $1/R$ potential in this case.

Although the  duality between $SU(N)$ gauge theory in four dimensions
 and type IIB string theory  expanded about an
$AdS_{5}\times {\bf S}^5$ background is  of great academic interest,
the most exciting extension of Maldacena's conjecture is to
non-supersymmetric gauge
theories,   especially to the physically relevant case of four-dimensional,
non-supersymmetric
gauge theory---namely  QCD$_4$. As Witten has shown \cite{w2},
it is reasonable to extend the conjecture to cases
where supersymmetry is broken, thereby obtaining properties of
non-supersymmetric gauge
theories in the large $N$ limit. For example, one can easily extend the
duality to discuss the
finite temperature behavior of the ${\cal N}=4$ gauge theory,  by
compactifying the (Euclidean)
time direction of the background space time of $AdS$ on a circle
of radius $\propto 1/{\bf temperature}$,
in which case supersymmetry is broken by the boundary conditions on the
circle. One can argue
that, since supersymmetry is broken, the fermions
and the scalars acquire a mass and, at least for large temperature
decouple, thus yielding a duality  to high temperature QCD.
Witten showed that in this case one derives  many of the expected features
of high temperature
gauge theory; including a non-zero expectation value of temporal (Polyakov)
loops, an area law
for spatial Wilson loops and a mass gap ({\it i.e.} a magnetic mass).

Finally, Witten has proposed a strategy  to study ordinary four dimensional
QCD at zero
temperature using string theory  \cite{w2}.
This can be done by using  Maldacena's conjecture to relate the large $N$
limit of the $SU(N)$-type ($2, 0$)  theory in ${\bf R}^6$  to
$M$ Theory on $AdS_7 \times S_4$ and dimensionally reducing these to four
dimensions.
(Throughout  this paper,
we consider theories on Euclidean signature spaces, and
$AdS_7$ here means its Euclideanized version.) To do this
and to break the
supersymmetry  one
sets the ($2,0$) theory on ${\bf S}^1 \times {\bf S}^1 \times {\bf R}^4$ with
supersymmetry breaking boundary condition on the fermions
around one of the ${\bf S}^1$'s.

An obvious candidate for its $M$ Theory dual
would be obtained by periodically identifying points on $AdS_7$
corresponding to the
periodicity's of the ${\bf S}^1 \times
{\bf S}^1$
and by imposing the supersymmetry breaking boundary condition
on the fermions by hand. There is, however, another candidate which obeys
the same
boundary condition. It is the Anti-de Sitter Schwarzschild solution
constructed by Hawking
and Page (for $AdS_4$ case) \cite{hp}. The supersymmetry breaking boundary
condition
is
automatically imposed by the Schwarzschild geometry. It
turned out that, the classical action for the
$AdS$ Schwarzschild solution is smaller than that of the vacuum $AdS_7$, and
therefore is dominant in the large $N$ limit \cite{w2}.

To make contact with four dimensional QCD  we must shrink the radii of the
two circles to
zero  in a certain limit. In  this construction, the six-dimensional
($2,0$) theory is regarded as a regularization of 
the four-dimensional QCD. The ultraviolet
cut-off scale is therefore set by the size of the compact space ${\bf S}^1
\times {\bf S}^1$.

Denote  the radius of the supersymmetry preserving
circle by $R_1$ and that of the supersymmetry breaking one by
$R_2$. The gauge coupling constant $g_{YM}$ of QCD$_4$ is given by
the ratio of the radii $g_{YM}^2 = R_1/R_2$. In the 't Hooft limit, where
one keeps
$g_{YM}^2 N$ to be finite,
the circle ${\bf S}_{R_1}^1$
shrinks to zero as one takes $N \rightarrow \infty$.
This corresponds to the IIA limit of $M$ Theory as ${\bf S}_{R_1}^1$ is the
supersymmetry
preserving circle. Therefore one could have started
with the theory on $N$ D$4$ branes in the IIA theory, wrapped around a
circle with
nonsupersymmetric  boundary conditions, rather than the six-dimensional
theory.  We will
take this approach  throughout  the paper. QCD is then regarded as the
dimensional reduction
of the five dimensional theory at high temperature, with coupling
$g_{YM}^2=g_5^2T$, where $g_5$ is the five dimensional  coupling
and $T$ the temperature (inverse radius) of the circle.

Witten has argued  that Wilson loops exhibit a confining area law behavior
in this geometry
for large $N$ and large  $g_{YM}^2 N$.
However, as he points out, this does not establish that QCD
is a confining theory.  The gauge theory so constructed   has an
ultraviolet cutoff
($\propto T$) and the coupling  $g_{YM}$ should be thought of as the bare
coupling
at distances corresponding to $1/T$. The string tension will turn out,  for
large
$\lambda = g_{YM}^2N$ (as we shall show below), to be proportional to $
\lambda
T^2$. To construct four dimensional QCD we must take
\be
T\to \infty     \ \ \ \  {\rm and } \ \ \ \
\lambda \to {b \over \ln\left({T\over \Lambda_{\rm QCD}}\right)},
\ee
where  $\Lambda_{\rm QCD}$ is the QCD mass scale.
Presumably we would find, were we able to calculate the small $\lambda$
behavior
of the tension,  that the tension
	behaves as  $ \exp[-{2b\over \lambda}]T^2
\sim \Lambda_{\rm QCD}^2.$ This calculation is beyond our control at the
moment. For
small $\lambda$
the background geometry develops singular behavior
and  the supergravity approximation surely breaks down. To deal with this
continuum
limit one would have to be able to calculate the properties of
string theory with background Ramond-Ramond charge
in a rather singular background.

Thus,  for the time being, the Maldacena-Witten  conjecture  only informs
us about the
behavior of large $N$ QCD, with a fixed ultraviolet
cutoff in the strong coupling(large $\lambda$)  regime. The resulting
physics should be
compared  best with strong coupling lattice gauge theory, where the lattice
spacing $a$ is
analogous to $1/T$, the radius of the fifth dimension. What is remarkable
here is that the
short distance cutoff, unlike in the case of lattice, 
does not destroy the
rotational or Lorentz
symmetry of the theory. Indeed, at short distances we see
a higher dimensional theory with more symmetry, indeed enough symmetry to
render the
theory finite. We are using  the
six dimensional, ultraviolet finite, (2,0) theory
to define the theory in the ultraviolet, yet its infrared behavior should
be qualitatively the
same as QCD.

In this paper we shall explore some of the features of  large $N$
supersymmetric and nonsupersymmetric gauge
theories using the above duality conjectures. We  shall show that the
resulting strong
coupling behavior of the gauge theories
is, in all cases, consistent with our qualitative expectations of
these theories.  Some of these consistency checks are highly nontrivial and
give additional
evidence for the validity of the proposed dualities.

First we shall explore, in the next section, the connected correlation
function of Wilson loops.
This kind of calculation can be used for  many purposes among which are the
evaluation of
the electric mass (or screening length) of high temperature QCD, the glue
ball spectrum of
confining gauge theories and the demonstration that in the confining phase
of QCD
monopoles are condensed. In particular we
outline how the  glueball spectrum of  this version of strong
coupling QCD could be calculated.

In Section III, we generalize the discussion of QCD to the case where the
$\theta$ parameter
is non-zero and  argue that we
can demonstrate oblique confinement.

In Section IV we generalize  the evaluation of Wilson loops in the
fundamental representation
to higher representations. Here we find that the string theory naturally
produces the behavior
of higher representations that 
we would expect in a confining theory---a result
that depends critically on the master field being described by fermionic
strings.

In Section V we argue that one can also use the duality to discuss
heavy quark baryonic states and determine the effective energy of $N$
fundamental
representation  quarks  in a singlet state
for large $N$. The construction of the baryon is possible because of
the Chern-Simons term in the action for supergravity on $AdS$.
The same arguments allow us to show that the interaction energy between 
any finite number of 
quarks is zero for the conformally invariant supersymmetric four
dimensional gauge theory  and infinite for the confining theory.

Finally, we conclude with a discussion of the possibility
of a large $N$ phase transition.  If  such a phase transition exists the
power of the
conjectured duality would be significantly weaker.

While this paper was being typed, we learned of the work
\cite{w3} where a similar construction of baryons is given.

\section{Confinement, Monopole Condensation and Glueball}

In this section, we first review the works \cite{w2,bisy,rty,li} where 
it was
shown how confinement in  strong coupling
QCD$_p$ can be seen  in the dual description
based on  $AdS$ supergravity. In particular,  
they demonstrated the area law behavior of the
Wilson loop expectation value. We then
discuss implications of this result and
clarify
an issue that was raised in \cite{bisy,rty,li} on the apparent
divergence of the electric and magnetic masses. It turns out
that this is related to the
computation of the mass gap suggested in \cite{w2}. We discuss
how one can compute glueball masses in this description.

According to Maldacena's conjecture \cite{malda1,imsy}, the
$(p+1)$-dimensional
maximally supersymmetric gauge theory realized as
the low energy dynamics of $N$ D$p$ branes ($p \leq 5$)
is dual to  type II string theory on the near horizon geometry of the D$p$
brane, as  given by
\be
l_s^{-2} ds^2 =
\sqrt{\frac{gN}{u^{7-p}}} du^2+
\sqrt{\frac{u^{7-p}}{gN}}\sum_{i=0}^p dx_i^2 +
\sqrt{gN u^{p-3}} d \Omega_{8-p}^2 ,
\label{susymetric}
\ee
where $l_s$ is the string length, $d \Omega_{8-p}$
is the line element of ${\bf S}^{8-p}$, and $g$ is related to the Yang-
Mills coupling constant. We have neglected numerical factors that are not
relevant to the following discussion. For $p \neq 3$, the dilaton $\phi$
depends on  $u$ and is given by
\be
e^{\phi} = g \left( \frac{gN}{u^{7-p}} \right)^{\frac{3-p}{4}} .
\label{dilaton}
\ee
In particular,
for $p=3$, the near horizon geometry (\ref{susymetric})
is $AdS$ and the dilaton (\ref{dilaton}) is constant,
corresponding to the
fact that
the theory on D$3$ brane is conformal. 
%Therefore, for large $gN$, we expect that the supergravity 
%description is appropriate.
%For $p>3$ ($p<3$), the
%curvature $l_s^2 R \sim \sqrt{u^{3-p}/gN}$
%becomes strong for small (large) $u$ and the dilaton blows
%up for large (small) $u$. There the supergravity description would fail,
%and we need
%to use the full $M$ theory description \cite{imsy}.

Witten \cite{w2} proposed to study non-supersymmetric QCD$_p$
by compactifying the supersymmetric theory in $(p+1)$ dimensions
on a circle and break the supersymmetry by imposing
anti-periodic boundary conditions on the fermions.
In the dual type II theory this  corresponds to considering
the $AdS$ Schwarzschild geometry
\bea
 l_s^{-2} ds^2 &=&\sqrt{\frac{gN} {u^{7-p}}}\frac{du^2}{1 -
u_0^{7-p}/u^{7-p}  } +	\sqrt{\frac{u^{7-p}}{gN}}\left(1 -
	u_0^{7-p}/u^{7-p}\right) d\tau^2 + \nn \\
& &+	\sqrt{\frac{u^{7-p}}{gN}}\sum_{i=1}^p dx_i^2
	+ \sqrt{gN u^{p-3}} d \Omega_{8-p}^2 ,
\label{geometry}
\eea
with the dilaton $\phi$  given by (\ref{dilaton}).
We can regard $(\tau,x_1,..,x_p)$ as coordinates for the
$(p+1)$-dimensional gauge theory.
To make the horizon at $u=u_0$
regular, the coordinate $\tau$ has to be periodically identified
as $\tau \rightarrow \tau + 1/T$ with $T$ being related to
$u_0$ by
\be
	u_0 = (gN T^2)^{\frac{1}{5-p}}.
\label{horizon}
\ee
Since the circle in the $\tau$-direction
is contractible at $u=u_0$, the boundary condition on the
fermions around the circle
is automatically anti-periodic, breaking the supersymmetry.
For large $T$, the $(p+1)$-dimensional theory becomes effectively
$p$-dimensional,
the fermions and scalars decouple, and the theory should resemble
QCD$_p$ in the
infra-red.

If QCD$_p$ is confining,  the vacuum expectation value of the Wilson loop
operator
$W(C)$ should exhibit area law behavior. In \cite{w2,bisy,rty}  this was
shown to
be the case, for large $gN$,  by
evaluating
the classical action of string worldsheet
bounded by a loop on ${\bf R}^p$ located at $u =\infty$. Because 
of the $u$-dependent factor $\sqrt{u^{7-p}/gN}$ in front of
$\sum_i dx_i^2$ in the metric (\ref{geometry}),
it is energetically favorable for the worldsheet to drop
near the horizon $u=u_0$  before spreading out in the ${\bf R}^p$
direction. At the horizon, the
$u$-dependent factor becomes
\be
\sqrt{\frac{u_0^{7-p}}{gN}}
=(gN)^{\frac{1}{5-p}}  T^{\frac{7-p}{5-p}},
\ee
where we used (\ref{horizon}).  Therefore the area dependent part of the
Wilson loop
expectation value becomes
\be
\langle W(C) \rangle = {\rm exp}(- (gN)^{\frac{1}{5-p}}  T^{\frac{7-p}{5-p}}
		A(C) ),
\label{wilsonvev}
\ee
where $A(C)$ is the area bounded by the loop $C$. Since the QCD$_p$
coupling constant
$g_{YM}$ is related to $g$ by $g_{YM}^2 = g T$, the string
tension derived from the above formula is
\be
	({\rm tension})_p = (g_{YM}^2 N)^{\frac{1}{5-p}} T^{\frac{6-p}{5-p}}.
\ee
For $p=3,4$, this agrees with the formulae derived in \cite{bisy,rty}.

\begin{figure}[htb]
\begin{center}
\epsfxsize=3.5in\leavevmode\epsfbox{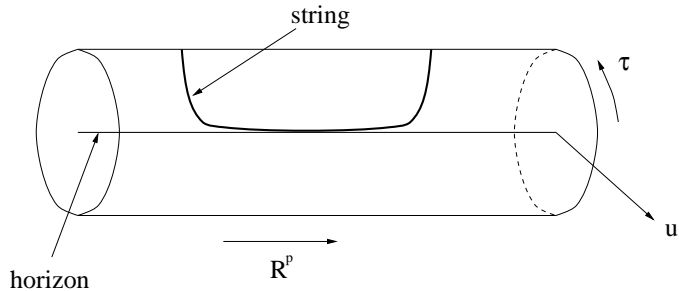}
\end{center}
\caption{The string drops to the horizon first before spreading
in the ${\bf R}^p$ direction.}
\label{fig3}
\end{figure}

In four dimensions, it is expected that  confinement is associated with
magnetic monopole
condensation. It is interesting to see that this in fact happens in this
construction \footnote{While
this work was in progress, we received \cite{li} where a related issue was
discussed.}.
To discuss QCD$_4$, we start
with the five-dimensional theory on D$4$ branes. The magnetic monopole
in five dimensions is a string which is realized as a D$2$ brane
ending on a D$4$ brane \cite{diaco,hw}. The monopole in four dimensions
is obtained by wrapping the string around the compactifying ${\bf S}^1$. It
is now
straightforward to compute the potential between
a monopole ($m$) and  an anti-monopole ($\bar{m}$).
Consider a pair of $m$ and $\bar{m}$
traveling along the $x_1$-axis in ${\bf R}^4$ and separated
in $x_2$ direction by distance $L$. In the large $gN$ limit, the force
between them is
mediated by a D$2$ brane bounded by ${\bf S}^1$ times
the trajectories of $m$ and $\bar{m}$, which are located at $u = \infty$.
Away from the
boundaries, the D$2$ brane can spread in the $u$-direction. In its
classical configuration,
$u$ would be a function of $x_2$ only because of the symmetry of the
problem. If we use
$(\tau,x_1,x_2)$
as the coordinates on the D$2$ brane, the induced metric on the brane is then
\bea
	G_{\tau,\tau}
	&=& \sqrt{\frac{u^3}{gN}}\left(1 - \frac{u_0^3}{u^3}\right) \nn \\
G_{11} &=&
\sqrt{\frac{u^3}{gN}}, ~~ G_{22} =
\sqrt{\frac{gN}{u^3}}\frac{(du/dx_2)^2}{1 - u_0^3/u^3}
						+ \sqrt{\frac{u^3}{gN}} .
\label{d2metric}
\eea
By taking into account the dilaton configuration (\ref{dilaton}),
which in this case is
\be
	e^{\phi} = g \left( \frac{u^3}{gN} \right)^{1/4} ,
\label{d2dilaton}
\ee
the D$2$ brane action per unit length in the $x_1$ direction
becomes
\bea
	E_{m\bar{m}} &=& \int_0^L 
d\tau dx_2 e^{-\phi} \sqrt{G_{\tau\tau}G_{11}G_{22}} \nn \\ 
&=& \frac{1}{gT} \int_0^L dx_2 \sqrt{ \left(\frac{du}{d
x_2}\right)^2
+ \frac{1}{gN}(u^3 - u_0^3)} ,
\label{d2action}
\eea
and it gives the potential energy for the $m$-$\bar{m}$ pair.

The next task would be to minimize this action. In fact,
essentially the same problem has already appeared in \cite{bisy,rty}
where  the correlation function of
temporal Wilson loops in five dimensions was studied.
 There one considers
a	string, rather than the D$2$ brane, wrapping in the $\tau$-direction
and spreading in the $x_2$-direction. Because of (\ref{d2metric}) 
and (\ref{d2dilaton}), we have
\be
e^{-\phi} \sqrt{G_{\tau\tau}G_{11}G_{22}} = \frac{1}{g}
\sqrt{G_{\tau\tau}G_{22}}.
\ee
Therefore the classical action of the string 
is equal to $g$ times
that of the D$2$ brane discussed in the above paragraph.
Therefore,
we can borrow the result of \cite{bisy,rty} to discuss the
$m$-$\bar{m}$ correlation.

The new feature of this problem is
a classical instability of the D$2$-brane worldvolume.
When the distance $L$ between $m$ and $\bar{m}$ is less
than a certain critical distance $L_{\rm crit}$, which is equal to
$1/T$ times some numerical factor, there is a D$2$ brane
configuration
minimizing the action (\ref{d2action})
and connecting $m$ and $\bar{m}$.
If $L$ exceeds this critical distance,
there is no connected D$2$ brane configuration minimizing
the action (Fig. 2). This happens because
the $G_{\tau\tau}$ component of the induced metric (\ref{d2metric}) can be
made arbitrary small by going near the horizon $u = u_0$ reflecting the
fact that
the compactification circle along $\tau$ is contractible in the $AdS$
Schwarzschild geometry. Since the
circle is contractible, the D$2$ brane can split into
two pieces
each of which has a topology of a disk and is bounded
by the trajectory of $m$ or $\bar{m}$. Therefore,
for $L > L_{\rm crit}$,
the potential between $m$ and $\bar{m}$ becomes
constant and the force between them vanishes.
This suggests that the magnetic monopole is completely screened. In this
construction, therefore,  confinement is in fact accompanied by  monopole
condensation. If we view this system as  finite temperature QCD$_{p+1}$,
such a complete screening indicates that the magnetic mass is infinite.
This is somewhat puzzling and we will address this issue later in this
section.

\begin{figure}[htb]
\begin{center}
\epsfxsize=3.5in\leavevmode\epsfbox{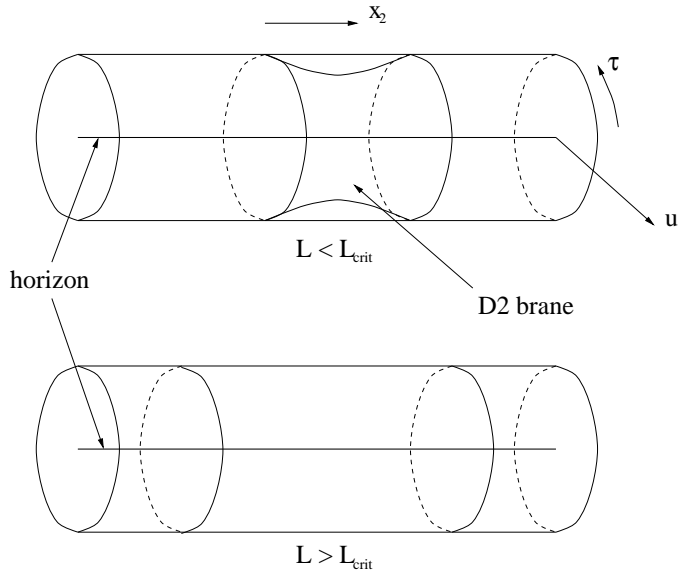}
\end{center}
\caption{For $L > L_{\rm crit}$, there is no volume-minimizing
D$2$ brane configuration connection the $m$-$\bar{m}$ pair}
\label{fig4}
\end{figure}

A similar classical instability also shows up when one studies correlation
functions of Wilson
loops. If one considers two Wilson loops in ${\bf R}^4$ and repeats  the
above analysis to
compute their correlation function,
one finds that, beyond a certain critical distance determined by the size
of the
loops, the correlation function vanishes identically. Once again, this is
because
the loops are contractible and the string stretched between the loops
becomes classically
unstable beyond the critical distance (Fig. 3). This result is again
somewhat puzzling
since one would
expect that the Wilson loops correlation for large distance would
be characterized by glueball exchange. This result seems to indicate that
the glueball mass
in QCD$_{p}$ is infinite. To address this issue, it is
useful to look into the nature of the classical instability and discuss
what happens at the
critical distance and beyond.

\begin{figure}[htb]
\begin{center}
\epsfxsize=3.5in\leavevmode\epsfbox{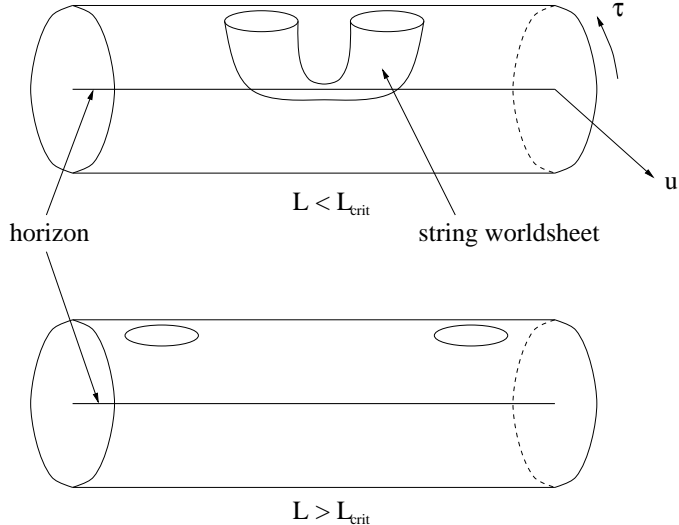}
\end{center}
\caption{For $L > L_{\rm crit}$, there is no area-minimizing
string worldsheet connecting the two Wilson loops. The critical distance
$L_{\rm crit}$ is
determined by the size of the loop.} \label{fig5}
\end{figure}

The instability of minimal surfaces has been know for a long time.
It was L. Euler who showed that a minimal surface bounded by
a two concentric circle in ${\bf R}^3$ is given by a catenoid.
Let us put the two circles of radius $R_0$ at $z =\pm L/2$. Euler's
catenoid is given by
\be
\sqrt{x^2 + y^2} = R_{\rm min} {\rm cosh}\left( \frac{z}{R_{\rm min} }
\right),
\label{euler}
\ee
where $R_{\rm min}$ is the minimum radius of the catenoid, which
is a function of the distance $L$ between the circles and the radius $R_0$
determined by the relation
\be
	R_0 = R_{\rm min} {\rm cosh}\left( \frac{L}{2R_{\rm min}} \right).
\label{eq1}
\ee
When the two loops coincide $(L=0$), obviously this formula gives
$R_{\rm min} = R_0$. As one increases $L$, the minimum radius $R_{\rm min}$
decreases. As shown in Fig. 4, however,
there is a critical value of $L_{\rm crit} = 1.325 R_0$. For 
$L > L_{\rm crit}$,
there is no
solution to (\ref{eq1}). There
the only minimal surface is a pair of disks bounded by the two
circles, called the Goldschmit discontinuous solution.
At $L = L_{\rm crit}$,
the catenoid becomes unstable. A small perturbation would make
the surface to pinch and split into the two disks.

\begin{figure}[htb]
\begin{center}
\epsfxsize=2.5in\leavevmode\epsfbox{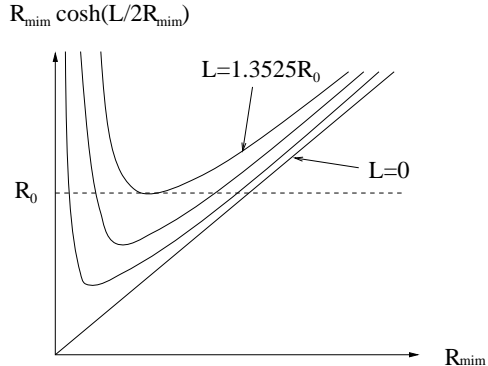}
\end{center}
\caption{For $0 < L < 1.3525 R_0$, the solid
curve $R = R_{\rm min} {\rm cosh}(L/2R_{\rm min})$ intersects
twice with the dotted line $R = R_0$,
determining the minimum radius
$R_{\rm min}$ of the catenoid. For $L > L_{\rm crit}$, there
is no intersection, indicating that a catenoid solution
does not exist.}
\label{fig0}
\end{figure}

At $L < \tilde{L}_{\rm crit} 
= 1.056 R_0$, the area of the catenoid is smaller than
that of the
Goldschmit solution and therefore the catenoid is absolutely stable. At $L
= \tilde{L}_{\rm crit}$,
the areas of the two solutions coincide and, for $\tilde{L}_{\rm crit}
 < L < L_{\rm crit}$,
the catenoid becomes more voluminous than
the Goldschmit solution. Therefore the transition from the catenoid to the
Goldschmit solution at $L = L_{\rm crit}$ is of the first order.

What does this mean for the Wilson loop correlation function? When the
distance between the loops $C_1$ and $C_2$ is less than the critical distance
$L< L_{\rm crit}$, the main contribution to the connected part of the
correlation function
$\langle W(C_1) W(C_2) \rangle$ comes from
the classical string connecting $C_1$ and $C_2$. At $L = L_{\rm crit}$, 
the string
worldsheet becomes unstable and starts to collapse.
Before the surface becomes disjoint, however, the supergravity approximation
breaks down when the radius of the cylinder becomes of the order of the string
length $l_s$. After that,
quantum fluctuations of the surface start to support
the worldsheet against the total collapse, and the two disks would be
connected by a
thin tube of a string scale $l_s$. For large $L$, the thin tube is
represented by the
supergraviton exchange between
the two disks. Therefore the correlation between the Wilson loops does not
completely
vanish, but are mediated by the supergraviton exchange between the disks
(Fig. 5).
This indicates
that the supergravitons in the $AdS_{p+2}$ Schwarzschild blackhole geometry
should
be identified with the glueballs of QCD$_p$.

\begin{figure}[htb]
\begin{center}
\epsfxsize=2.5in\leavevmode\epsfbox{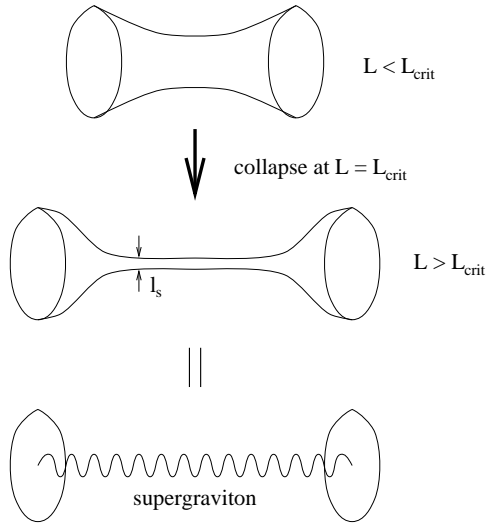}
\end{center}
\caption{The string worldsheet connecting the Wilson loops collapses
at $L = L_{\rm crit}$ 
and is replaced by the supergraviton exchange.} \label{fig1}
\end{figure}

Another way to obtain glueball masses would
be to compute correlation functions of local operators in
QCD$_p$ and look for particle poles. According to
\cite{gkp,w1}, a two-point correlation function of local
operators in
the $(p+1)$-dimensional supersymmetric gauge theory is
obtained by computing the Green's function of the
corresponding supergraviton (or its Kaluza-Klein cousin) on
$AdS_{p+2}$. Similarly
a	correlation function in QCD$_p$ should be
related to a Green's function on the $AdS_{p+2}$
Schwarzschild geometry. The glueball
masses computed in this way would be the same
as the one that appeared in the Wilson loop
correlators in the above paragraph as they
are both extracted from the supergraviton propagator.

Since the bulk geometry
is invariant under translation in the ${\bf S}^1 \times {\bf R}^p$
direction, we can expand the supergraviton wave $\phi(u,\tau,x)$
in the Fourier modes as
\be
\phi(u,\tau,x) = \sum_n \int {d^pk \over (2\pi)^p}
		\tilde{\phi}_{n,k}(u) e^{in\tau + ikx}.
\ee
Each Fourier mode corresponds to a particle
pole of the correlation function on ${\bf R}^p$
with mass $k^2$. Those with $n\neq 0$ are
Kaluza-Klein modes on ${\bf S}^1$ and are not of interest for
QCD$_p$. For a given $k^2$, the Fourier
mode $\tilde{\phi}_{n,k}(u)$ obeys the second order ordinary
differential equation for $u$. Witten showed
in \cite{w2} that the equation has a regular
solution only for discrete values of $k^2$, suggesting the mass
gap in QCD$_p$.

In order to actually compute
the glueball masses, one has to solve the differential equation. In
the $AdS_{p+2}$ geometry, the differential equation for the
supergraviton has three regular singularities and therefore can be
solved analytically using the hypergeometric function
\cite{earlier,gkp}. For the Schwarzschild
geometry, the differential equation has four regular
singularities (for QCD$_3$), with the additional singularity coming from the
horizon, and requires numerical work.
Various aspects of the glueball
spectrum are currently under study using this technique \cite{cortt}.

\begin{figure}[htb]
\begin{center}
\epsfxsize=3.5in\leavevmode\epsfbox{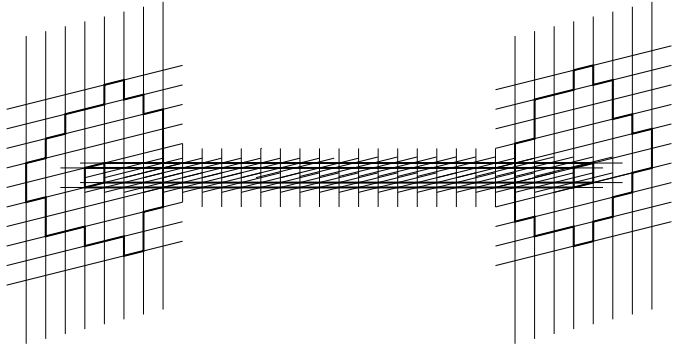}
\end{center}
\caption{The glueball in the strong coupling lattice QCD.} \label{fig6}
\end{figure}

The glueball masses computed in this way are
quantized in the unit of
the temperature $T$. One might have worried that in the effective
large $N$ string theory of glueballs the mass gap (the lowest mass
of glueballs) would vanish --- corresponding to the massless mode
of the string. In strong coupling lattice QCD,
the leading contribution to the Wilson loop correlator comes
from a thin rectangular
tube of the size of the minimum lattice spacing $a$, as shown in Fig. 6.
Therefore the glueball masses are of the
order of $1/a$ in the strong coupling. Indeed, with the standard Wilson
lattice action, the glueball mass, for strong coupling is given by $m_{\rm
glue}=4/a\log(g^2N)[1+O(1/g^2N)]$. To make contact with the real
world, one would have to sum the strong coupling expansion to obtain
$m_{\rm glue}=4/af(g^2N)$, and then let $g^2N\to b/\log(1/\Lambda_{QCD} a)$
as $a \to 0$, obtaining (hopefully) a finite result proportional to the QCD
scale $\Lambda_{QCD}$. 
In the $AdS$ picture $1/T$ plays the role of $a$, and as the
string shrinks to distances of order $1/T$ the fluctuations
in the extra dimensions produce a finite mass gap proportional
to $T$. Thus $m_{\rm glue} \propto T f(\lambda)$. 
As in the case of the string tension discussed above, the
computation of the $4d$ glueball spectrum would 
require control of the string theory in a singular background with R-R
charge.

\section{$\theta$ Parameter and Oblique Confinement}

\begin{figure}[htb]
\begin{center} \epsfxsize=4.5in\leavevmode\epsfbox{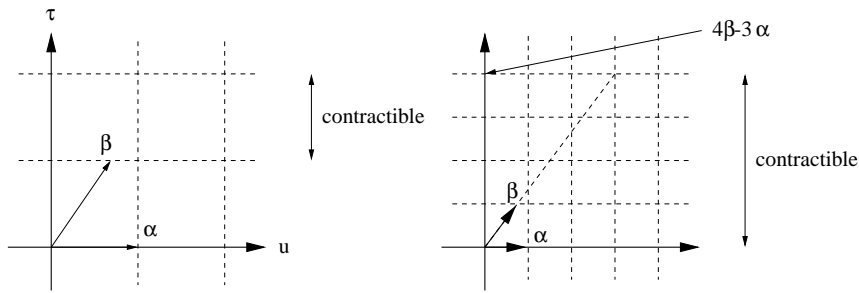}
\end{center}
\caption{When $\theta/2\pi$ is a rational number,
one can find a contractible cycle on the torus.}
\label{fig7}
\end{figure}

It is interesting to generalize the discussion of  the previous section to
the case of
QCD$_4$ with non-zero $\theta$ parameter.
Its $M$ Theory dual can be constructed as follows.
The $\theta$ parameter couples to $\int {\rm tr} F \wedge F$
in four dimensions, which is the D$0$ brane charge on the D$4$ brane.
Therefore, if
we view QCD$_4$ as the high
temperature theory of the theory in five dimensions, $\theta$ can be
interpreted as the
chemical potential for D$0$ branes. In $M$ Theory, this is geometrically
realized as a
rotation of the supersymmetric circle by $\theta$ as ones
goes around the supersymmetry breaking circle once.
As shown in \cite{w2}, the $M$ Theory dual of
QCD$_4$ with $\theta= 0$ is the $AdS_7$ Schwarzschild
solution times ${\bf S}^4$, given by
\bea
N^{-2/3} ds^2 &= & \left(u^2  - 1/u^4 \right)
				d\tau^2
					+ g_{YM}^2 u^2 d \rho^2 + \nn \\
 && + \frac{du^2}{u^2 - 1/u^4} + u^2 \sum_i dx_i^2
			+ d \Omega_4^2,
\eea
with periodicities
\bea
\alpha-{\rm cycle}:~~(\tau,\rho) &\rightarrow& (\tau,\rho+2\pi) \nn\\
\beta-{\rm
cycle}:~~(\tau,\rho) &\rightarrow& (\tau+2\pi,\rho) .
\eea
We can turn on $\theta$ in this geometry
by introducing a twist
as
\bea
		\alpha-{\rm cycle}:~~(\tau,\rho) &\rightarrow&
(\tau,\rho+2\pi) \nn \\
			\beta-{\rm cycle}:~~(\tau,\rho) &\rightarrow&
(\tau+2\pi,\rho+\theta).
\eea

For non-zero $\theta$, the $\beta$-cycle is not contractible.
Correspondingly the monopole
condensation does not take place.
In fact the $m$-$\bar{m}$ potential would obey the area law
in this case.
If $\theta = 2 \pi p/q$
for some co-prime integers $(p,q)$, however,
the cycle $(q \beta - p \alpha)$ becomes
contractible in the Schwarzschild geometry (Fig. 7). A membrane ending on
this cycle gives a
dyon of an electric charge $-p$ and a magnetic charge $q$, and this dyon is
screened since the
membrane worldvolume can collapse. In this case,  confinement is associated
with the
condensation of
this $(-p,q)$ dyon, corresponding to oblique confinement.

\section{Higher Representations}

Another interesting generalization is  to consider Wilson loops for higher
representations
of $SU(N)$. Using  the Frobenius formula, \be
	\chi_R(U) = \frac{1}{n!} \sum_{\sigma \in S_n}
\chi_R(\sigma) \prod_{i=1}^{K_\sigma}
{\rm tr}~ U^{k_i},
\label{frobenius}
\ee
one can relate the character $\chi_R(U)$ of a representation $R$ to a
product of
traces in the fundamental representation of $SU(N)$ as expressed in the right-
hand side of the equation.
Here $n$ is the number of boxes of  the Young tableau for $R$,
$\chi_R(\sigma)$
is the character of the permutation $\sigma$
in the representation of the symmetric group
$S_n$ associated to the same Young tableau. A permutation $\sigma$
can be expressed as a product of cycles, and $k_1,...,k_{K_\sigma}$
are lengths of the cycles. Therefore,
the Wilson loops expectation values for higher representations are related to
multiply wound loops in the fundamental representation. The latter are
computable by studying string worldsheet ending
on such loops.

Let us summarize what we expect for the Wilson loops for
higher representations from the field theory analysis. In QCD$_2$,
they are all computed exactly \cite{gt}
as
\be
\langle \chi_R(U) \rangle = ({\rm dim} R) {\rm exp}(-g_{YM}^2 C_2(R) A(C)),
\label{higher}
\ee
 where $C_2(R)$ is the quadratic Casimir of $R$ given by
\be
	C_2(R) = n N + \sum_i r_i^2 - \sum_i c_i^2 - \frac{n^2}{N},
\label{casimir}
\ee
with $r_i$ and $c_i$ being the lengths of the rows and the columns
of the Young tableau for $R$.

It would be instructive to look at some examples. For $n=2$, there
are symmetric ($S$) and anti-symmetric ($A$) representations.
Their quadratic Casimirs are
\bea
		C_2(S) =& 2N + 2 - \frac{4}{N} \nn \\
		C_2(A) =& 2N - 2 - \frac{4}{N}.
\eea
The Frobenius formula (\ref{frobenius}) gives
\bea
\chi_S(U) =& \frac{1}{2}[ ({\rm tr} U)^2 + {\rm tr} U^2 ]\nn \\ \chi_A(U)
= &\frac{1}{2}[ ({\rm tr} U)^2 - {\rm tr} U^2],
\eea
which can be inverted as
\bea
\langle  {\rm tr} U^2 \rangle &=& \chi_S(U) - \chi_A(U) \nn \\ &=& N^2 e^{-2
g_{YN}^2 N A(C)}
\times \left( - 4g_{YM}^2 A + \frac{1}{N} + \cdots \right).
\eea
In general, (\ref{higher}) combined with the Frobenius formula gives \be
\frac{1}{N^n} \langle   {\rm tr} U^n \rangle =
(-1)^{n-1} g_{YM}^{2n-2} \left( A^{n-1} + O\left(\frac{1}{g_{YM}^2 N}
\right) \right)
			{\rm exp}(-ng_{YM}^2 N A(C)) .
\label{sign}
\ee
The factor $(-1)^{n-1} A^{n-1}$ is closely related to the presence of
the quadratic Casimir in the string tension and it is reasonable to expect
that this factor would
appear in other confining theories as well.

The prefactor in (\ref{sign}) has the form of the $\alpha'$ expansion
of the string theory. One might hope therefore to be able to derive the
leading term $(-1)^{n-1} A^{n-1}$ without detailed knowledge of
string theory on $AdS_{p+2}$. Let us see how far we can go.
To evaluate $\langle {\rm tr} U^n \rangle $ using Maldacena's duality, one
sets the
Wilson loop $C$ at $u=\infty$ and considers
a	string worldsheet whose boundary winds around $C$ $n$-times.
With this boundary condition, the minimal surface in the $AdS_{p+1}$
Schwarzschild
geometry (\ref{geometry}) should look as follows.
As in the case of $n=1$ in section 2,
because of the $u$-dependent factor
in front of $\sum_i dx_i^2$, the worldsheet quickly drops to
the horizon. After that,  the worldsheet is allowed to spread in the ${\bf
R}^p$
direction. For fixed $u$,
the metric in the ${\bf R}^p$ direction is
 flat, so that we can use flat space intuition.
It is possible to construct a smooth surface whose boundary winds
around a circle $n$-times, but one can always reduce an area of such a
surface by shrinking a part of the surface and creating branch points.
The minimal surface constructed in this way should then  have a
form of $n$-disks on
the top of each other connected by $(n-1)$ $Z_2$-branch points \footnote{In
${\bf R}^3$,
it is known that a minimal surface bounded by any regular curve is smooth
without branch point \cite{oss}. The boundary contour in our problem does
not satisfy an assumption of this theorem
as it winds around the identical circle $C$ $n$-times.}. The classical
action for the
minimal surface is then
\be
S_{n-winding} = n (gN)^{\frac{1}{5-p}}  T^{\frac{7-p}{5-p}}
		A(C).
\ee

To understand the prefactor $(-1)^{n-1} A^{n-1}$, we have to
go beyond the supergravity approximation. In general,  stringy corrections
are difficult to control since we know little about the formulation of string
theory in a background with Ramond-Ramond (R-R) charges. Fortunately,
 we only need to study the $0$-th order in
the $\alpha'$-expansion to understand the prefactor, consequently
we can neglect the R-R charges
as well as the spacetime curvature. Therefore,  we can treat
the string worldsheet as described by the free fields of the NS/R string.
All we have to do then is to compute the disk amplitude of free string
theory with $(n-1)$ branch points. To the $0$-th order in the $\alpha'$-
expansion, this simply amounts to computing determinants of the
Laplace operators
on the disk with $(n-1)$ branch points and summing over all possible locations of
the branch points.
The factor $A^{n-1}$ is then easy to explain as it corresponds to the entropy
factor associated to
the positions of the branch points.

The sign factor $(-1)^{n-1}$ is more interesting. It cannot come from the
determinant factor  since  the worldsheet does not have obvious unstable
modes. We claim that it is a consequence of the GSO projection of the
superstring theory. To the $0$-th
order in $\alpha'$, the R-R background does not affect the worldsheet
theory and the
distinction between NS and R states is well defined.

When $n=1$ (no branch point), 
the fermions around the contour $C$ obey    NS-NS
boundary conditions. Let us remind ourselves why this is the case. As a
function of a
coordinate $z$ on the disk
($|z| \leq 1$), the fermion $\psi(z)$ in this case is
single valued since there is no singularity in the interior
of the disk. To study the boundary condition around the contour
$C$ at $|z|=1$, it is appropriate to use another coordinate $\theta$
defined by $z =
e^{i\theta}$.
Since the fermion is a spin-1/2 field, we have to multiply
the transition function
$\sqrt{dz} = e^{\frac{i}{2} \theta} \sqrt{d \theta}$. This means that a
fermion
obeys  anti-periodic boundary conditions around $C$, i.e. it is in the
NS-NS sector.

For $n >1$, we must  take into account the presence
of the branch points. Near $|z|=1$, we can use the covering coordinate $w$
which is
related to $z$ by $z = w^n$. To change
  coordinates  from $z$ to $w$, we have to multiply the transition function
$\sqrt{dz} = w^{\frac{n-1}{2}} \sqrt{dw}$. To study the boundary
condition around $C$, we use
the coordinate $\theta$ defined by $w = e^{i \theta}$, and multiply the
transition function
$	w^{\frac{n-1}{2}} \sqrt{dw} = e^{i \frac{n}{2} \theta}
\sqrt{d\theta}$ to the fermions.
It is then clear that a fermion will  obey   anti-periodic or periodic
boundary conditions around $C$ depending on whether $n$ is odd or even.
Therefore,  the closed string emitted from the Wilson loop $C$ is in the
NS-NS or in the R-R sector depending on the parity of $n$.

It is known that the GSO projection requires that amplitudes in the R-R
sector in this case should be multiplied by the sign factor $(-1)$ relative to
that in the NS-NS sector. This was observed, for example, in \cite{pol} and
was found to be responsible for the cancellation between the exchange
forces of the NS-NS and the R-R fields. The string worldsheet in the
$AdS$ Schwarzschild geometry has a quasi-cylindrical region where the
worldsheet quickly drops from $u=\infty$ to the horizon $u=u_0$ to save
the energy. If we look at this region in the open string channel, to
reproduce the fermion boundary condition for even $n$,
one has to insert the fermion number parity operator $(-1)^F$ as one goes
around the loop $C$. It is then clear
that the GSO projection
require the sign factor $(-1)$ for the corresponding amplitude. It is
interesting to note that this result depends critically on the fact that
the master field is described by
fermionic string. This confirms a long-standing conjecture as to the 
fermionic
nature of the large $N$ 
string theory that describes a confining gauge theory \cite{migdal,gt}.

So far we have discussed the confining case, but
the same argument should be applicable to the conformal case as well, leading
to the sign factor $(-1)^{n-1}$
in $\langle {\rm tr} U^n \rangle$. In order to
see whether this is what one naturally expects, it is useful
to first point out some puzzling feature of Maldacena's computation of the
Wilson loop in the conformal case \cite{malda2}. For a rectangular Wilson
loop of sides $L$ and $R$ ($L \gg R$) in the ${\cal N}=4$
theory in four dimensions, he finds
\be
	{1 \over N} \langle {\rm tr} U \rangle
= {\rm exp}\left( + {4 \pi^2 \over \Gamma(1/4)^4}
{\sqrt{2 g_{YM}^2 N} L \over R} \right)
\label{ineq}
\ee
for the fundamental representation.
Since $U$ is a unitary matrix, there is an upper bound on its expectation
value
${1 \over N} \langle {\rm tr} U \rangle \leq 1$ which contradicts with the
sign
in the exponent
in the right-hand side of (\ref{ineq}).
One possible resolution of this puzzle would be that
the unitarity bound is violated due to a renormalization of the operator $U$.
The renormalization of $U$ corresponds
to the mass renormalization of the quark going around the loop, and
we expect it to be zero in the ${\cal N}=4$ theory. However the
nonrenormalization theorem
assumes a supersymmetric regularization, which would
typically violate the inequality of this type. With an explicit
ultraviolet
cutoff, $\Lambda$, which may break the supersymmetry but preserves the
inequality, the Wilson loop expectation value would be \be
{1 \over N}  \langle {\rm tr} U \rangle
= {\rm exp} \left( - c L \Lambda   + {4 \pi^2 \over \Gamma(1/4)^4}
{\sqrt{2 g_{YM}^2 N} L \over R} \right)
\ee
for some positive constant $c$. If the mass renormalization $c \Lambda$ for a
representation $R$ is proportional to its quadratic Casimir, which is
reasonable, one would find
\be
 \langle {\rm tr} U^n \rangle
 \sim (-1)^{n-1} \left( L^{n-1} + \cdots \right)
		{\rm exp}\left(- nN c_0 L\Lambda +
 {4 \pi^2 n \over \Gamma(1/4)^4}
{\sqrt{2 g_{YM}^2 N} L \over R} \right),
\ee
obtaining the sign factor $(-1)^{n-1}$ again. The factor $L^{n-1}$,
as opposed to $A^{n-1}$ in the confining case, indicates that the twist
operators are constrained
to stay along the minimum $u$ point in the string worldsheet.
It would be interesting to understand this phenomenon better
from the point of view of the string theory in the $AdS$
background.

\section{Baryons and Stringy Exclusion Principle}

The $M$ Theory dual of QCD$_4$ can also be used to construct hadrons
with heavy quarks and study their properties. To obtain mesons,
one simply starts with a pair of quark and anti-quark represented by Wilson
lines of
opposite orientations
on a D$4$ brane separated from $N$ D$4$ branes.  The quark anti-quark pair
is then
connected by a string extended through the $AdS$ geometry
(the Schwarzschild solution or $AdS$ depending on whether one
study the pure
QCD or the superconformal theory). The string generates a potential
between the
quarks, which is either linear (confining case) or is inversely
proportional to the distance
between the quarks (conformal case). One can then study the
non-relativistic quantum
mechanics of the quarks in this potential to compute
the meson spectrum. This is essentially the same analysis as the Bag model.

\begin{figure}[htb]
\begin{center}
\epsfxsize=3.5in\leavevmode\epsfbox{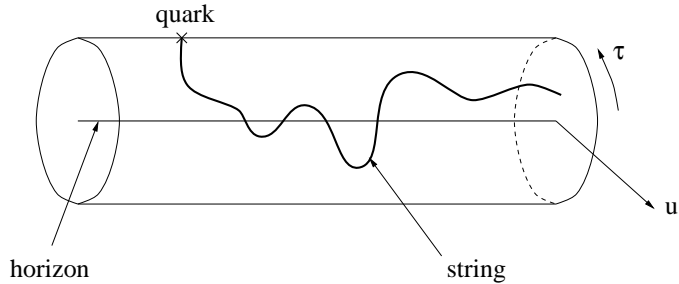}
\end{center}
\caption{In the Schwarzschild geometry, there is no point
where the open string attached to the quark can end.}
\label{fig8}
\end{figure}

There is no free quark of finite energy in the confining case; whereas in
the conformal
case finite mass colored states exist.  How do we see this in the
supergravity picture?
Both are   a consequence of the fact that the string attached to the quark
must  end
somewhere. In the conformal case, the string can end at $u=0$. The energy
of such a
string is simply equal
to the BPS mass of the quark \cite{malda2}.
In the confining case, the bulk geometry has the
 topology of a solid cylinder
with the horizon
$u=u_0$ as its axis as shown in Fig. 8.
In this case there is no point where the open string can end
and the string has to stretch to $x = \infty$ in the
${\bf R}^p$ direction, costing
an infinite amount of energy.

This raises the interesting question of how one could construct
a baryon,   a $SU(N)$ 
singlet bound state of $N$ quarks, in this picture.  We
have to find a way to tie together
$N$ strings emerging  from the quarks. In the flat space this is not possible
because of the conservation of the NS-NS two-form charge. A quark and
an anti-quark can be connected by a string since the NS-NS charge is
canceled at the two end points. It is not obvious how this can be done with
$N$ end points with the same charges.
 It is an amazing
consequence of supergravity
that it is possible to do so in the $AdS$ space.
In the conformal case, the dual supergravity on $AdS_5$ has the
Chern-Simons term for the
$SL(2,Z)$ doublet two-forms $B^a_{\mu\nu}$ ($a=NSNS, RR$)
\be
S_{CS_5} = {N \over 2}\epsilon_{ab} \int_{AdS_5} B^a \wedge d B^b.
\label{CSforB}
\ee
This comes from the fact that the equation of motion for $B^a$
in ten dimensions contains a coupling to the 5-form \cite{john} as
\be
D^\mu \partial_{[\mu,} B^a_{\nu\rho]}
 = - g \epsilon_{ab} F_{\mu\nu}^{~~~\rho\sigma\kappa}
		\partial_\rho B^b_{\sigma\kappa}.
\label{john}
\ee
A Lagrangian density
which gives such an equation of motion would be\footnote{
Although the complete Lagrangian for the IIB supergravity
in ten dimensions is not known, for the purpose of the discussion here, we
only
need a Lagrangian for the two-form.
Since we are only considering the classical supergravity,
any Lagrangian which gives (\ref{john}) should be good enough.} \be
{\cal L}_{B} = \frac{1}{2g} |dB^a|^2  +
{\epsilon_{ab} \over 2} F \wedge B^a \wedge dB^b. \label{iib}
\ee
On $AdS_5 \times {\bf S}^5$, the 5-form $F$ carries
$N$ units of flux on ${\bf S}^5$. Thus, for $B^a$'s which are
constant on ${\bf S}^5$, the ten-dimensional Lagrangian (\ref{iib})
implies the Chern-Simons
term (\ref{CSforB}) in five dimensions. Similarly the $3$-form
$C_{\mu\nu\rho}$ in the eleven-dimensional supergravity has the Chern-
Simons term
$	\int C \wedge dC \wedge dC $, which
upon compactification on ${\bf S}^4$, gives the seven-dimensional Chern-
Simons term on $AdS_7$
\be
	S_{CS_7} = \frac{N}{2} \int_{AdS_7} C \wedge d C.
\label{CSforC}
\ee
As we now explain, these Chern-Simons terms make it possible for the $N$
strings to
combine together and end on
a point in the bulk $AdS$ geometry.

Before explaining why the $N$ strings can join together 
in $AdS$, it would be
instructive
to review a similar but more familiar phenomenon in the three-dimensional
gauge
theory with the Chern-Simons term,
\be
	S_{CS_3} = \frac{N}{2} \int AdA
\label{CSforA}
\ee
for an abelian gauge field $A$. Suppose the electric charge of the theory
is quantized so
that we allow a gauge transformation $A \rightarrow A
+	d \theta$ with $\theta$ defined modulo integer. Let us perform a gauge
transformation by $\theta$ which has a discontinuity $\delta \theta = 1$
across a two-dimensional surface 
$D_2$ with the topology of disk. The variation of the
Chern-Simons
term under this gauge transformation is \bea
	\delta S_{CS_3} &=& N \int_{AdS_3} d(\theta dA) \nn \\
		&=& N \int_{D_2} dA = N \oint_{C} A,
\eea
where $C$ is the boundary of the disk $D_2$.
Thus, if there is a Wilson loop $C$ carrying $N$ units of charge,
we can absorb it into the Chern-Simons term by the gauge transformation. For the same
reason, if we put $N$ particles of unit charge on the
top of each other,
the composite particle decouples from the gauge field $A$.

The above story can be immediately generalized to the case of
string with the Chern-Simons term (\ref{CSforB}). The supergravity action is
invariant under the gauge transformation $B_{\mu\nu} \rightarrow B_{\mu\nu}
+ \partial_{[\mu,} \lambda_{\nu]}$. Since the $B$-field charges are quantized
with the unit charges carried by the fundamental string and the D string,
the gauge transformation parameter
$\lambda$ does not have to be a single-valued vector field on
$AdS_5$, but its integral $\oint_{C} \lambda$ around a closed contour $C$
can jump by
an integer amount. Suppose such a discontinuity
of $\lambda$ occurs across a four-dimensional subspace $M_4$ of $AdS_5$.
Since the
discontinuity $\delta \lambda$ has to be such that $\oint_{C} \delta
\lambda$ is an integer
and therefore is invariant under smooth deformation of the contour $C$,
$\delta \lambda$
is closed $d \delta \lambda = 0$ on $M_4$ and it can locally be
written as $\delta \lambda = d \theta$ with $\theta$ is defined modulo
integers.  Under
such a gauge transformation, the
Chern-Simons term (\ref{CSforB}) changes as
\bea
\delta S_{CS_5} &=& N \epsilon_{ab} \int_{AdS_5}  d(\lambda^a d B^b) \nn \\
 &=& N \epsilon_{ab} \int_{M_4} \delta\lambda^a dB^b
=	N \epsilon_{ab} \int_{M_4} d(\theta^a dB^b).
\eea
Suppose further that $M_4$ has the topology of ${\bf S}^1 \times M_3$ with
$M_3$ being a
three-dimensional space bounded by a two-dimensional surface $\Sigma$ and that
$\theta^{(RR)}$ jumps by
$1$ across a point on ${\bf S}^1$ (times $M_3$) while $\theta^{(NSNS)}$ is
continuous.
The gauge variation of $S_{CS_5}$ can then be written as
an integral of $B$ over $\Sigma$,
\be
 \delta S_{CS_5} = N \int_\Sigma B^{(NSNS)}.
\ee
Therefore $N$ string worldsheets wrapping on a contractible surface
$\Sigma$ in $AdS_5$ can be absorbed into the Chern-Simons term
(\ref{CSforB}) by the gauge transformation. This also means that, if we
put $N$ strings on the top of each other, it decouples from $B^{(NSNS)}$.

The string can be viewed as a soliton of the supergravity with the
$B^{(NSNS)}$-charge \cite{dhgr}.
The fact that $N$ strings 
decouple from $B^{(NSNS)}$ suggests that one can construct
a supergravity solution
in which $N$ strings join together at a point in $AdS_5$ 
\footnote{While this paper was being typed, 
we learned of the work \cite{w3} where it is shown that $N$
strings can end on the 5-brane wrapping on ${\bf S}^5$ and localized on
$AdS_5$. This verifies our claim that such a supergravity solution should
exist in $AdS_5$.}.
This defines a baryon in the conformal
case. The baryon in QCD$_3$ is obtained by simply compactifying this
picture on
${\bf S}^1$. Because of the Chern-Simons term (\ref{CSforB}), $N$ strings can
end on a point in the Schwarzschild geometry. The resulting string
configuration
is very similar to the one suggested much earlier in \cite{w4}.

It is obvious that this phenomenon also holds for closed strings ---
$N$ closed strings
can join together and disappear in $AdS_5$. This may be viewed as a higher
dimensional generalization of
the stringy exclusion principle
pointed out by Maldacena and Strominger \cite{ms}.
They showed that there is an upper bound
on the number of BPS particles in $AdS_3$. Although they derived this
result by
studying the spectrum of chiral primary fields
in the dual conformal field theory in two dimensions, it is also
possible to show this using the Chern-Simons term for the $SU(2) \times
SU(2)$ gauge field
in $AdS_3$ \cite{mar}.
The mechanism we described in the
above is a natural generalization of this to $AdS_5$ with
the gauge field being replaced by $B^{(NSNS)}$.

Let us turn to the case of QCD$_4$.
To construct baryons in this case, we have to start with a membrane
of the supergravity in $AdS_7$. Wrapping the membrane on the supersymmetry
preserving
circle gives a string on the $AdS_7$ Schwarzschild geometry. The membrane
carries a
charge with respect to the 3-form $C$, and the Chern-Simons term
(\ref{CSforC}) can
create a membrane with
$N$ units of charges. Therefore, once again, $N$ open strings can join
together in the bulk.

\begin{figure}[htb]
\begin{center}
\epsfxsize=3.5in\leavevmode\epsfbox{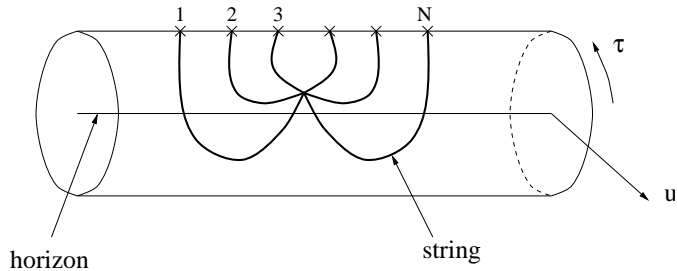}
\end{center}
\caption{Baryon is constructed as $N$ strings joining together
at a point in the bulk.}
\label{fig9}
\end{figure}

Thus, in both the conformal and confining cases, a baryon
is constructed as $N$ heavy quarks joined together by $N$ open strings
ending at a
point in the $AdS$ geometry.
The total energy of the baryon is the sum of the geodesic
length of the strings. In the confining case, it is proportional to a sum
of four-dimensional distances between the quarks
and the location of the string junction (projected to ${\bf R}^4$). In the
conformal
case, it is a sum of (distance)$^{-1}$. Although the location of
the string junction is dynamical, in the large $N$ limit,  we
can use the Born-Oppenheimer approximation and regard it as a fixed point
in the
$AdS$ geometry. The $N$ quarks then move independently under the potential
given by the string stretched between them and the junction. The mass
spectrum of
the baryon can them be
obtained by solving a one-body problem of the quark in the potential.

\begin{figure}[htb]
\begin{center}
\epsfxsize=3.5in\leavevmode\epsfbox{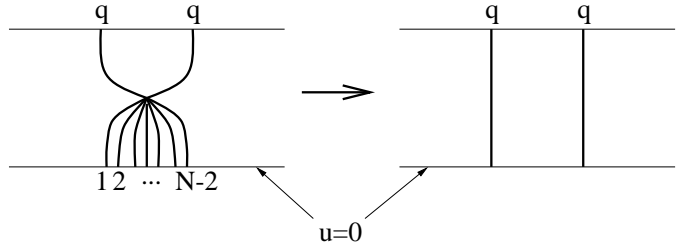}
\end{center}
\caption{It is energetically favorable to move the
$N$ string junction to $u=0$, leaving the two separate strings. }
\label{fig2}
\end{figure}

It would perhaps be worthwhile to point out that the existence
of the $N$ string junction does not contradict with the vanishing
of the $q$-$q$ potential in the conformal case. If you have a $q$-$q$ pair,
in addition
to the obvious string configuration
where a string stretched from each quark ends on $u=0$, one may consider
joining
them at a point from which $(N-2)$ strings come out. This may seem to give
a non-trivial potential 
between the quarks. However the $(N-2)$ strings should end
somewhere, and the only place they can end is at $u=0$. For $N >4$, it is
clear that
it costs less energy if we move the $N$ string junction toward $u=0$. We
then end
up recovering the obvious configuration
where the two strings separately end at $u=0$, without a non-trivial potential
between $q$-$q$ (Fig. 10).

A similar configuration can be considered in the confining
case. In this case, however, there is no place where the
$(N-2)$ strings can end, except at $x=\infty$ in ${\bf R}^p$ costing
infinite energy to create (Fig. 11), as expected from the quark
confinement. The same discussion holds for any non-singlet
combination of quarks.

\begin{figure}[htb]
\begin{center} \epsfxsize=2.5in\leavevmode\epsfbox{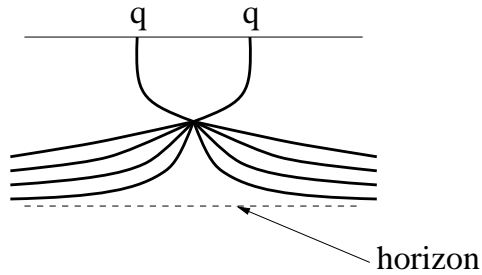}
\end{center}
\caption{A similar configuration in the confining
case costs infinite energy, as expected from the
quark confinement. }
\label{fig10}
\end{figure}
\section{Discussion}

In the previous sections we have
discussed  many aspects of the dynamics of large $N$ gauge theory
for strong coupling, as calculated using the dual
supergravity, string or $M$ theory. In all cases the results
are remarkably consistent with our intuition and expectations.
This strengthens our faith in the validity of the duality between these
pictures.
However, there is no strong evidence
to date that these pictures actually overlap ---
they could be descriptions of two quite
different phases of the same theory.

Consider the duality  between  the conformally invariant ${\cal N} =4$ gauge
theory
in four dimensions and string theory in the $AdS_5 \times {\bf S}^5$ 
background. The
strongest form of this duality is to claim that these two are equivalent for
all values of N and $\lambda$. However, perturbative string theory is
defined as an
asymptotic expansion in $g_s=g^2_{\rm YM}=\lambda/N$. So perhaps a safer
conjecture
is the  equivalence of these two formulations for $N=\infty$,
the classical limit of string theory and the planar limit of the gauge
theory, or in the
asymptotic $1/N$ expansion.

Here one is  on firmer ground since the weak coupling
expansion of  the $N=\infty$ conformally invariant gauge theory might very
well
converge \cite{thooft,banks} and there is no reason to expect the $1/\lambda
\sim {\rm curvature}$ expansion of  string theory to diverge.
If so, one could then imagine, in principle,
using the large $\lambda$ expansion (or even better the
exact solution)  of classical string theory to define
the gauge theory for all coupling. If one could do this for the
compactified  non-supersymmetric theory as well,
then one could construct continuum QCD, by taking $\lambda$ to
zero  \`a la asymptotic
freedom, as one lets the compactification radii vanish.
So far there is no strong evidence for even this form of the
conjecture;
since the calculations of quantities in string theory can only be done for
large
$\lambda$ and in the gauge theory
only for small $\lambda$.

There is, however, a possibility that the conjecture has to be weakened even
further. Namely, it is possible  that the
gauge theory picture is valid for weak coupling, the string theory for strong
coupling, and there is no region where they are both
valid --- {\it i.e.} there is a  phase transition at $\lambda=
\lambda_{\rm cr}$. This is a ubiquitous phenomenon
in large $N$ theories \cite{bg},
including the one-plaquette model \cite{gw},
QCD$_2$ on the sphere \cite{dk,gm}
 and is believed to be the case (even for finite $N$)  for lattice QCD. In
the case
of lattice QCD  such a transition
leads one to suspect that the effective string theory that can be deduced
from the
strong coupling
expansion  cannot be extended to the continuum theory.
If this were the case here as well, it would mean, unfortunately,
that these dualities are not as powerful as one might hope,
and in particular one might not be able to use them
to construct the master field, string theory, of QCD.
On the other hand,  the existence of such a phase transition
would make the conjectured duality seem more reasonable,
by eliminating  some of the  paradoxical aspects
of the duality \cite{vafa}.

\bigskip
\bigskip
\bigskip

\centerline{\bf Acknowledgments}

\bigskip

We would like to thank Tom Banks, Gary Horowitz, Emil Martinec, 
Yaron Oz, John Schwarz and many other participants 
of ITP Program, {\it Duality in String Theory}, 
for useful discussions.
The work of DG is supported in part by NSF grants NSF-PHY-94-07194
and NSF-PHY-97-22022.
The work of HO is supported in part by NSF grant NSF-PHY-95-14797 and
by DOE grant DE-AC03-76SF00098, and also by
NSF grant NSF-PHY-94-07194.

\end{document}